# Towards the Formation of a Positronium Coherent Beam

M. Sacerdoti[1], V. Toso[2,3], G. Vinelli[4,5,6], G. Rosi[4,5], L. Salvi[4,5,6], G.M. Tino[4,5,6],

M. Giammarchi[3], and R. Ferragut[1,3]

[1] L-NESS and Department of Physics, Politecnico di Milano, Via Anzani 42, 22100 Como, Italy

[2] Department of Physics, Università degli Studi di Milano, Via Celoria 16, 20133 Milano, Italy

[3] INFN, Sezione di Milano, Via Celoria 16, 20133 Milano, Italy

[4] Department of Physics and Astronomy, Università di Firenze, Via Sansone 1, 50019 Sesto Fiorentino, Italy

[5] INFN, Sezione di Firenze, Via Sansone 1, 50019 Sesto Fiorentino, Italy

[6] LENS, Via Nello Carrara 1, 50019 Sesto Fiorentino, Italy

**Abstract.** Positronium (Ps) has emerged as a promising test particle within the QUPLAS collaboration for investigating the gravitational effect. In this work, we present a novel approach to generate a monoenergetic and highly coherent Ps beam by creating a negative Ps ion (Ps-, consisting of two electrons and one positron). The necessary positron beam is formed by using a high flux electron LINAC. Subsequently, we utilize a Fabry-Perot IR laser cavity operating at a wavelength of 1560 nm to selectively remove the extra electron. An alternative pulsed laser operating at a 3600 nm wavelength was studied to reduce broadening due to recoil and excitation. Here, we provide a Monte Carlo simulation to estimate the characteristics of the Ps beam, including its energy distribution and intensity profiles. The results obtained from this study will provide essential groundwork for future advancements in fundamental studies as Ps gravity measurements by using a Mach-Zehnder interferometer.

## 1. Introduction

Gravitational studies with antimatter systems are of paramount importance to address Physics at the Planck scale, the current uppermost conceptual limit of validity of our physical theories. Quantum Gravity effect might in fact induce violation of the two basic symmetries that relate Quantum Mechanics and General Relativity: the Einstein Equivalence Principle (EEP) and the CPT (Charge-Parity-Time) invariance. The most complete theoretical framework that includes the possibility of Lorentz violation (and therefore EEP and CPT violation) is the Standard Model Extension (SME) [1,2]. The SME includes all the Physics known in the form of the Standard Model and the General Relativity, while adding to the Lagrangian the possible spacetime operators violating Lorentz invariance. In this frame, research with positronium if particularly interesting, since (contrary to the antihydrogen case) positronium is a fundamental system composed by fermionic mass alone - while the antiproton mass is mostly due to the energy of the QCD color field. In addition, the coefficients for Lorentz invariance violation are flavor-dependent [3]; for these reasons positronium gravitation constitutes an entirely new and independent window on Physics beyond the validity of our current best theories. For instance, in the frame of the SME, it has been shown that gravitation with Ps can address relevant physical parameters already at the 10% accuracy level [1].

Positronium (Ps) is a bound system of an electron and a positron and exists in the ground state in two sub-levels, according to the spins of the two particles: singlet called para-Ps with spin 0 and a lifetime of 0.125 ns and triplet called ortho-Ps with spin 1 and a lifetime of 142 ns. Para-Ps annihilates with the emission of two gamma rays of 511 keV each while ortho-Ps emits at least three gamma rays with a maximum energy of 511 keV each. Due to the short lifetime of para-Ps only ortho-Ps can be used in the experiment and Ps will always correspond to ortho-Ps in this paper for practical reasons.

This work presents a proposal for generating a monoenergetic and highly coherent Ps beam through the creation of a negative Ps⁻ ion [7]. The advantage of producing Ps through negative Ps⁻ ions, in contrast to Ps formed in meso-porous materials [4-6], is the ability to achieve a tunable and monoenergetic beam with low divergence and high coherence, making it well-suited for interferometric experiments. However, the drawback is the short Ps⁻ lifetime (479 ps), necessitating the use of a very compact system to efficiently produce, accelerate, and focus the Ps⁻ beam using electrodes. Subsequently, a laser system selectively removes the extra electron to form the Ps beam.

The obtained results lay essential groundwork for future advancements in fundamental studies, including Ps gravity measurements using a Large Momentum Transfer Mach-Zehnder interferometer [8].

## 2. Results and discussion

An effective positronium negative ion (Ps⁻) producing target consists of a tungsten (W) coated with a sodium (Na) sub-monolayer, as demonstrated in the last decade by Nagashima's group [7].

A Monte Carlo simulation was performed to quantify the Ps beam characteristics as a function of various optimization factors as Ps⁻ formation, guiding, and photo-detachment. For this purpose, we introduced a possible optimized velocity distribution for Ps⁻ emission, a miniaturized focusing and accelerating optics, and two possible laser photo-detachment systems.

## 2.1 Ps⁻ emission velocity distribution

The velocity distribution of the Ps⁻ emitted from the target surface is hereby proposed. In general, there are many potential effects that could influence the velocity distribution of the emitted ions from the target surface with a negative work function (see for instance Ref. 9). The thermal effect contribution given by the temperature of the emitter target is one of the most important consequences. Another effect that impacts the emission, especially visible in monocrystals, is the periodic lattice structure of the target and its electron energy distribution [10]. In fact, in the case of photoelectron emission, the angle-resolved photoemission spectroscopy (ARPES) technique to study crystalline solids is based on this principle. There is also evidence of this effect for the positronium emission with negative work function from aluminum (Al) single crystals [11]. Another effect, probably more relevant in semiconductors, is the excitation of electron-hole pairs. In fact, it could influence the scattering angles recently observed in semiconductors [12], which are higher than those observed in metals [13].

On the other hand, it is crucial to consider the surface topology and contamination that affect the directional emission of particles, often not considered in the literature. From the theoretical point of view the negative work function contributes with an emission component perpendicular to the surface. Therefore, the surface roughness increases the broadening of the angular distribution of the emission. For example, if only an effect of thermal contribution is expected for Ps in metals (see Refs. 13 and 14), the combination of this effect with that of surface roughness could be interpreted as a broadening of the angular distribution of emitted particles, which would appear to be emitted at higher temperatures.

In particular, the negative ion Ps⁻ emission was not previously studied in detail. An approach based on experimental results obtained in W and Al for positronium and positrons [13-15] emitted with negative work function is hereby proposed. Following this experimental evidence and considering "ideal" conditions such as a flat surface of the W target (coated with Na) the velocity distribution is influenced by the negative work function and the thermal effect, in the case of particles that reach the equilibrium with the metal phonons, considering the hypothesis of a negligible scattering effect. We are aware that this is an idealized condition for the calculations and that the actual velocity distribution is likely to be wider. However, this approach not only serves as a guide for determining the optimal conditions to maximize ion emission but also enables the projection of an optimized optic to guide the beam, as proposed in the next section (Sec. 2.2).

In the directions parallel to the surface of the W target (called $x$ and $y$-directions or transversal direction, see Fig. 1a), the contribution will result in the thermal effect. A classical two-dimensional

Maxwell-Boltzmann distribution is proposed in this case, where the pre-exponential factor depends on the first power of the velocity.

$$f(v_T) = 2\,\alpha\,v\,exp(\alpha v^2), \tag{1}$$

where $\alpha$ depends on the target temperature $T$,

$$\alpha(T) = \frac{1}{2}\frac{m_{Ps^-}}{kT} = \frac{3}{2}\frac{m_0}{kT}, \tag{2}$$

$m_{Ps^-}$ and $m_0$ are the negative ion Ps$^-$ and electron masses and $k$ the Boltzmann constant.

The problem of particle emission in the direction perpendicular to the surface (called the $z$ or longitudinal direction, see Fig. 1 (a)) with a negative work function is rarely addressed in the literature. However, some works deal with the emission of positrons with a negative work function to obtain energy distributions with experimental results [16]. They claim that the particles are directed likewise a collimated beam of gas molecules effusing as in the Stern-Gerlach experiment, following a Maxwell-Boltzmann distribution. The proposed velocity distribution takes the following form,

$$g(v_L) = \beta \left(\frac{1}{2}mv^2 - \phi\right)^{3/2} \exp(-\alpha v^2), \tag{3}$$

Note that the exponent applied to the parenthesis is 3/2 to follow the pre-exponential tendency of $v^3$ proposed in Ref. 16. Additionally, the negative work function $\phi$ is considered explicitly as the main contribution of the Ps$^-$ ion initial kinetic energy in this longitudinal direction. The minimum Ps$^-$ kinetic energy is valid by definition if higher than the absolute value of $\phi$. To obtain $\beta(T)$ it was necessary to normalize Eq. 3 integrating by using a numerical method at different temperatures. Rearranging Eq. 3 to be implemented in the Monte Carlo simulation,

$$g(v_L) = c_z \alpha^2 \left(\frac{\phi}{kT}\right)^{1/2} (v^2 - v_0^2)^{3/2} \exp[-\alpha(v^2 - v_0^2)] = c_z \alpha^2 \left(\frac{\phi}{kT}\right)^{1/2} \delta v^3 \exp(-\alpha \delta v^2), \tag{4}$$

where

$$v_0^2 = \frac{2}{3}\frac{\phi}{m_0}, \tag{5}$$

and

$$\delta v^2 = v^2 - v_0^2. \tag{6}$$

Equation 4 is defined zero if $v$ is less than $v_0$. The constant $c_z$ in eq. (4) results $\approx 3/2$. The absolute value of the negative work function for Ps$^-$ is estimated to be $\phi \approx 4$ eV when utilizing a poly-crystalline W target coated with a Na sub-monolayer, or a mono-crystalline [100]-oriented W target.

It is noteworthy to underline that the measured absolute value of the negative work function for Ps⁻ in these materials is approximately 1 eV [7]. The presence of the Na coating effectively reduces the negative work function by approximately 3 eV (see Refs. [7] and [17]).

## 2.2 Ps⁻ acceleration and focus optics

Investigation of the initial conditions is crucial to optimize the optical configuration needed to guide and focus the ions to form a coherent Ps⁻ beam. Initially, a monoenergetic positron beam is implanted in a sodium-coated tungsten (W) sample. This sample operates in transmission geometry, serving as a positron/Ps⁻ converter. As a result, a cylindrical beam of negative (Ps⁻) ions with a diameter of 1 mm is generated. Although Ps⁻ ions have a limited lifetime of 479 ps, they have the advantage of being influenced by Coulomb forces, allowing efficient guiding and focusing. The emission of Ps⁻ ions occurs perpendicular to the converter surface, with a kinetic energy of ≈4 eV as introduced in Sec. 2.1. A thermal contribution is added according to the sample temperature and following the velocity distribution proposed before.

An electrostatic system with electrodes at different electrical potentials has been developed. The short lifetime of the Ps⁻ requires the development of an extremely compact electrode design at a sub-millimetric scale. To realize this electrostatic system and reconstruct the Ps⁻ beam trajectory, a simulation was developed using the SIMION® software. This software calculates the trajectories of charged particles in a region of space where the electric fields are generated. These fields are calculated from geometrically defined electrodes with a set of applied potentials.

This simulation was developed considering the following goals: (i) accelerate the Ps⁻ ions to a final kinetic energy of 300 eV; (ii) focus the beam with a minimum divergence given the initial conditions; (iii) develop a compact electrodes design to minimize the Ps⁻ annihilation. The electrode geometry, designed based on these criteria, is shown in Fig. 1. In particular, Fig. 1 (a) gives a schematic representation, not drawn to scale. For a more accurate representation, Fig. 1 (b) presents the scaled scheme used for the SIMION® simulation.

The system consists of a positron/Ps⁻ converter (a monocrystalline Na-coated W sample), a focusing ring, and a grounded electrode with a hole centered in the beam axis. The converter has been idealized as a disk with a radius of 2.5 mm to which a potential of -300 V is applied. This is repulsive for the Ps⁻ ions which are therefore accelerated towards the grounded electrode to reach 300 eV kinetic energy. A 1 mm diameter hole has been opened in the center of the last electrode to allow the beam to pass through. To avoid propagating electrostatic fields outside this system, this hole has also been closed with a conductive grounded grid. The high transmission grid consists of a series of metal wires (1 μm in diameter) with a pitch of 50 μm.

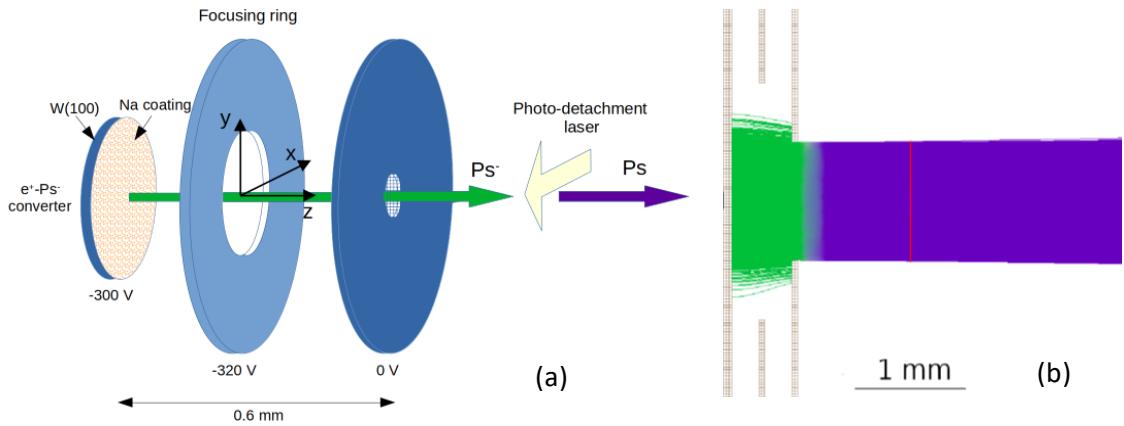

**Figure 1.** (a) Schematic representation of the optics, not drawn to scale. (b) SIMION® simulated beam represented to scale. The green trajectories represent the Ps⁻ ion beam, while the violet trajectories represent the Ps beam after photo-detachment. Both Ps⁻ and Ps trajectories are plotted solely for visualization purposes, without considering their natural decay. The green/violet transition lines represent the photo-detachment interaction zone. The vertical red line corresponds to the initial position of the excitation and Mach-Zehnder interferometer stages (see Ref. 8).

The distance between this last electrode and the converter is 0.6 mm. This ensures the dielectric strength necessary to avoid electric discharges given the applied potential differences. To better control and reduce the divergence of the beam, a focusing ring with a 2 mm diameter was also inserted in an intermediate position between the two electrodes just described (0.3 mm after the converter). The potential to be applied to this electrode in order to minimize the divergence of the beam is -320 V.

Ps⁻ are generated by the SIMION® simulation from the surface of the converter. Their initial position $(x, y)$ is randomly extracted within a Gaussian distribution centered with the center of the converter and with a standard deviation of 0.2 mm. This simulation provides the electric field configuration necessary for the Monte Carlo simulations.

### 2.3 Photo-detachment cavity and designed interferometer

Immediately after the last electrode, the Ps⁻ beam is hit by a laser beam propagating along the *x*-axis to produce a Ps beam via photo-detachment. Two lasers at two different wavelengths were studied for this process: 1560 nm and 3600 nm. The shorter laser wavelength was chosen to be very close to the peak of the photo-detachment cross section [7], while 3600 nm wavelength was studied to reduce broadening due to recoil and excitation.

The 1560 nm CW laser system consists of an erbium fiber laser pumped into a high finesse Fabri-Perot enhancement cavity to accumulate a circulating power of about 200 kW. The ultra-low absorption mirrors and setup are similar to those used in the PVLAS experiment [18]. Instead, the 3600 nm laser will be a pulsed laser with similar characteristics to Ref. 19.

A Mach-Zehnder interferometer was designed to perform Ps gravity measurements (see Ref. 8). To complete a gravity measurement in a reasonable time, the interferometer requires a high number of Ps atoms with angles between ±125 μrad and ±170 μrad in *y* and *x*-directions respectively. The Mach-Zehnder interferometer, located after the optics described in this work, measures the gravitational acceleration through the phase shift induced by a set of light pulses to the Ps wavefunctions in a gravitational field. Compared with a moiré deflectometer, this interferometer has the advantage of no grating absorption.

After the photo-detachment cavity at 1.6 mm from the converter, Ps atoms are first excited from the $1^3S_1$ level to the $2^3P_0$ level by a 243 nm laser then to the $2^3S_1$ level by an 18.25 GHz circularly polarized microwave radiation. A similar UV-microwave excitation scheme can be found in Ref. 20. The $2^3S^1$ state has a lifetime of 1.14 μs and allows for relatively long propagation through a Mach-Zehnder interferometer. The atomic wave-function is split in two branches and the internal Ps state oscillates between the $2^3S_1$ and $3^3P_2$ states due to the interaction with several laser pulses having a frequency equal to 1312.2 nm. At the interferometer output one of the branches is removed with a laser ionization stage of level $3^3P_2$ and two electrodes remove the remaining positrons and electrons to prevent them from contributing to the detected noise. Since the number of atoms exiting one of the two interferometer arms is proportional to the phase shift, it is possible to measure the gravitational acceleration of Ps simply by means of a detector which counts the impinging atoms. The following steps after the photo-detachment are described in detail in Ref. 8, where a simulation of the time needed to measure the gravity acceleration with a precision of 10% is also presented.

## 2.4 The Monte Carlo simulation

A Monte Carlo simulation was conducted to calculate the trajectories of $10^6$ Ps⁻ ions in the electric field configuration obtained from SIMION®. The simulation traced the path of the ions from the converter to the entrance of the Mach-Zehnder interferometer. The path of each ion is computed in every time step (tenth of picosecond), linearly interpolating the field.

With the aim of comparing the dependence of the final Ps spatial distribution on the velocity distribution at two different Ps⁻ emitting targets temperatures, simulations at room temperature (300 K) and cryogenic temperature (10 K) were performed.

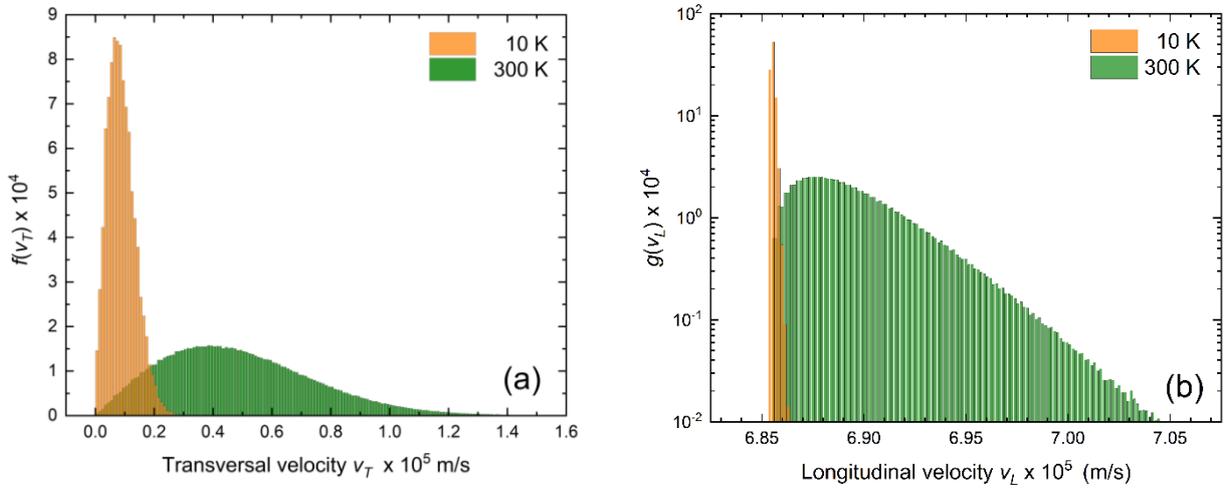

**Figure 2.** (a) Transversal velocity distribution in the *x* or *y*-directions. (b) Longitudinal velocity distribution in the *z*-direction. Both distributions are obtained at two different Ps⁻ emitting targets (10 K and 300 K) by Monte Carlo simulations following the proposed Eqs. 1 and 4, respectively. The $g(v_L)$ distribution is presented in logarithm scale to improve the visualization.

The Ps⁻ are generated from a disc surface of 1 mm in diameter following a Gaussian distribution. The velocity and direction of Ps⁻ are obtained adding the three-velocity components $v_x$, $v_y$ and $v_z$ following the distributions proposed in Sec. 2.1 (Eqs. 1 and 4). The realistic conditions that $g(v_L)$ in the *z*-direction tends to zero when the velocity *v* is less than $v_0$ or greater than $v_0$ plus half the thermal velocity $v_{th}$, which is given by $v_{th} = \alpha^{-1/2}$ (where $\alpha$ is given by Eq. 2), were imposed for the simulation. The thermal velocity $v_{th}$ results ~$10^4$ m/s at 10 K and ~$5.48 \times 10^4$ m/s at 300 K. Instead, the distribution of the transversal velocity $f(v_T)$ tends to zero when the velocity *v* is greater than 5 times the thermal velocity. The upper limits for the simulated distributions are higher than the natural limit of the proposed functions given by Eqs. 1 and 4. Figure 2 displays the velocity distributions of $10^6$ Ps⁻ ions obtained at both 10 K and 300 K. Each figure panel illustrates the dependence of the distributions on temperature: if the temperature decreases, the distributions become sharper and the average velocity decreases. These findings reveal the optimum initial conditions for achieving a monochromatic Ps beam before acceleration (Sec. 2.2). On the other hand, the initial divergence of the beam at low temperature, given by the angular beam aperture $\theta \approx \frac{2v_T}{v_L}$, will be reduced and localized if compared with high room temperature condition.

When the Ps⁻ exit from the last electrodes grid, the code simulates the photo-detachment performed by a laser beam along the *y*-axis. The probability of detachment at each time step is given by:

$$P_{ph} = 1 - exp\left(\frac{\lambda \sigma}{hc P_1 P_2} \int_{t_1}^{t_2} \int_{P_1}^{P_2} I(s,t) ds dt\right), \tag{7}$$

where $\lambda$ is the laser wavelength, $\sigma$ is the photo-detachment cross section, $h$ is the Planck constant, $c$ is the speed of light and $I(s,t)$ is the laser intensity. Assuming that the optical cavity is along the *x*-direction and that the laser can be seen by the atomic beam as a standing wave in both the continuous and pulsed regime (long pulses), the intensity takes the form

$$I(x,y,z,t) = 4I_0 exp\left(\frac{-2(y^2+z^2)}{w_0^2}\right) cos^2(kx), \tag{8}$$

where $w_0$ is the minimum laser beam waist.

The integral of intensity is numerically evaluated between the positions of the Ps in its trajectory each time step according to its velocity.

When the Ps⁻ is photo-detached, the recoil of the Ps, due to the absorption of the photon and the one due to the detachment of the electron, are simulated. The absorption of the photon induces a recoil along the laser direction (*x*-axis) with a positive or negative sign. Given the low energy of the photon, this recoil is negligible compared to the effect of electron detachment.

The two bodies separation recoil depends on the residual energy after breaking the bond between one of the electrons and the newly formed positronium atom and consequently on the wavelength of the photon. This energy is given by the difference between the photon energy of the laser and the photo-detachment threshold which is equal to 0.32688 eV [19].

We call $\varphi$ the angle between the velocity variation caused by the detachment recoil and the *z*-axis, and $\omega$ the angle between the *x-y* plane velocity projection and the y-axis. By setting the electric field of the laser along the *z*-axis, $\varphi$ has a "cosine-squared" distribution [21,22] so that the most probable values are nearest to zero while the distribution of $\omega$ is uniform in all directions.

One way to avoid this effect is to use a laser having energy close to the photo-detachment threshold energy, which corresponds to a wavelength of about 3800 nm. In this region the cross-section is about one order of magnitude smaller than the maximum with a laser of 1560 nm [7]; this implies that a laser power higher than 200 kW is required to keep the photo-detachment efficiency high. A 3600 nm pulsed laser system can reach 2 MW with high repetition rate [23] and can then replace the CW 1560 nm laser. Using the 3600 nm laser, the effect of photo-detachment on beam divergence is minimized.

The plots in Fig. 3 represent the Ps angular distributions $\theta_{XZ} = tg\left(\frac{v_x}{v_z}\right)$ for Ps$^-$ emitting targets at T=10K and T=300K. The distribution in the perpendicular direction $\theta_{YZ}$ is almost identical to $\theta_{XZ}$, the possible differences are within computational error. The dominant effect of recoil in this scenario is the electron detachment. Since this effect occurs in a random direction, it equally impacts the $\theta_{YZ}$ and $\theta_{XZ}$ distributions, resulting in a similar heating effect on the Ps beam.

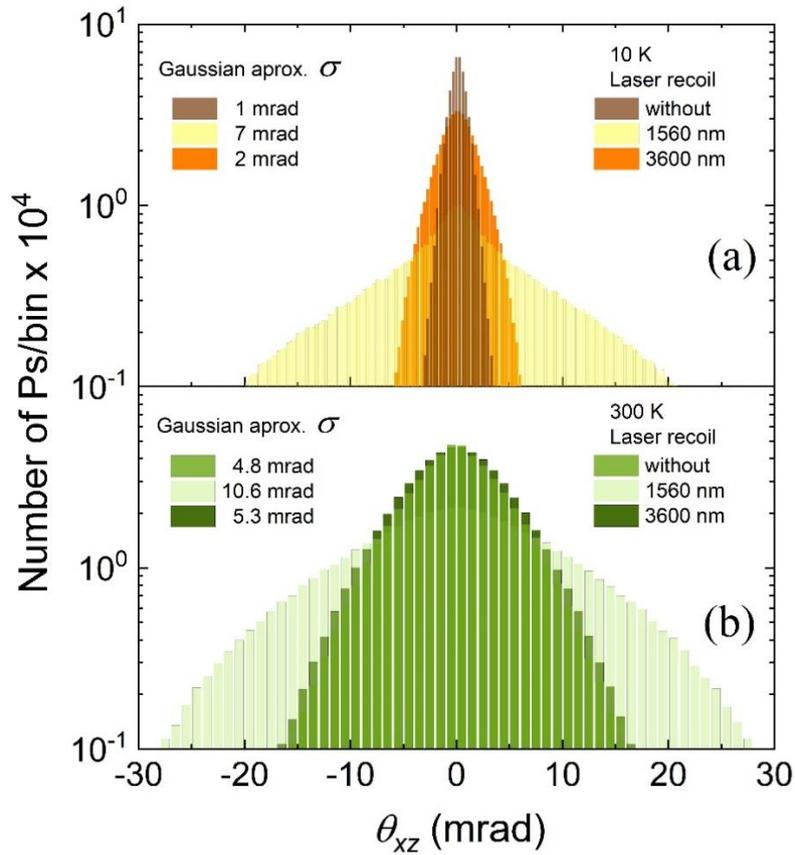

**Figure 3.** Angular distributions $\theta_{XZ}$ of Ps after laser recoil during photo-detachment obtained via Monte Carlo simulations. The angular distributions $\theta_{YZ}$ are almost identical to the $\theta_{XZ}$, the possible differences are within computational error. Results obtained for two different Ps$^-$ emitting targets temperatures: at 10 K in (a) and at 300 K in (b).

A hypothetical recoilless distribution for an "ideal" laser is also included in Fig. 3 to visualize the electron detachment effect in the Ps beam. The distributions follow a Gaussian-like trend. On the left of Fig. 3 (a) and (b) the approximate standard deviation $\sigma$ of the distributions (affected with an error +/- 0.2 mrad) is shown to compare the influence of the laser recoil on the width of each profile as a function of the wavelength (1560 nm and 3600 nm). The distributions for Ps¯ emitting targets at 10 K and 300 K follow a similar pattern. The 3600 nm wavelength minimizes the laser recoil effect, especially when the Ps¯ emitting target is at room temperature (300 K).

With this setup, the effect of recoil is to broaden the angular distribution of the Ps beam mainly in the $z$-direction, while minimizing its divergence in the $x$ and $y$ directions. This choice is particularly convenient since the interferometer is more sensitive in $x$ and $y$ broadening than $z$-direction.

Figure 4 shows the spread in kinetic energy in the $z$-direction after the laser recoil during photo-detachment. The results demonstrate that the Ps energy spread mainly depends on the laser wavelength, the influence of the Ps¯ emitting targets temperature is almost negligible in the $z$-direction. The net effect in this direction is to produce a symmetric spread in energy of about +/- 12%, when the 1560 nm laser is used, instead the spread is minimized up to approximately +/- 2% using the 3600 nm laser for the photo-detachment. Following the simulations presented in Ref. 8 the $z$ broadening effect reduces only few percent of the interferometer efficiency and results in being non-destructive.

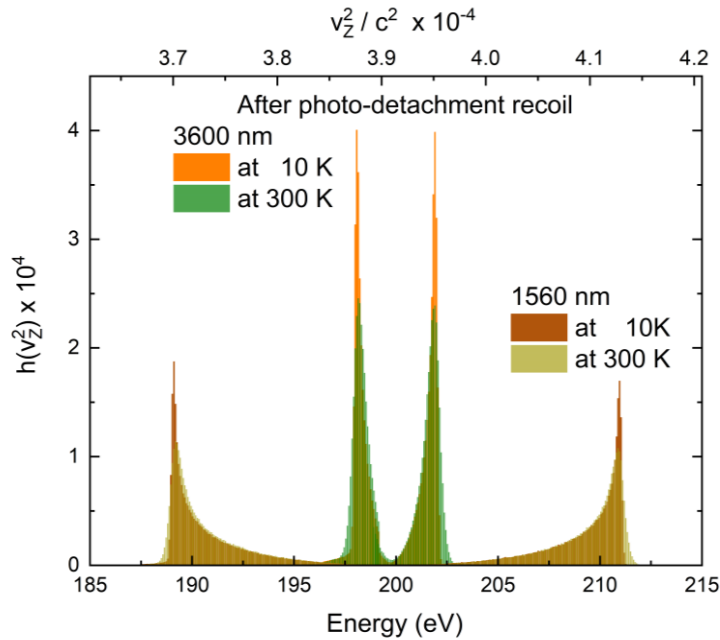

**Figure 4.** Ps energy distribution after the photo-detachment recoil in $z$-direction. The distributions are symmetric compared to the Ps average kinetic energy (200 eV). The recoil effect is simulated for two different laser photo-detachment wavelengths, 1560 nm and 3600 nm, and two Ps¯ emitting targets temperatures, 10 K and 300 K.

To make a comparison, both the lasers at 1560 nm and 3600 nm have been simulated. After crossing the laser beam, the positions and velocity components of the atoms that survived annihilation are stored and analyzed. Only the Ps in a 0.4×0.4 cm² square area are considered. The distance between the converter and the interferometer entrance is 1.6 mm (see Fig. 1 (b)). The results of the simulation for the laser wavelength 1560 nm and 3600 nm and for two Ps⁻ emitting targets temperatures (10 K and 300 K) are shown in Table I. The results indicate that after the emission of $10^6$ Ps⁻, the conversion efficiency Ps/Ps⁻ is around 57% and around 55% in average reach effectively the Mach-Zehnder interferometer. Now the more relevant limitation is the divergency tolerance of the interferometer of the order of a hundred of microradians. The last two rows of Table I show the efficiency in the *x* and *y*-direction. The better efficiencies were achieved by using the 3600 nm laser, especially when the Ps⁻ emitting target is at cryogenic temperature.

| Laser wavelength | 1560 nm | 1560 nm | 3600 nm | 3600 nm |
| --- | --- | --- | --- | --- |
| Temperature | 10 K | 300 K | 10 K | 300 K |
| Ps⁻ decayed before laser | 326,475 | 326,309 | 326,475 | 326,309 |
| Ps⁻ decayed in the laser | 92,343 | 85,753 | 68,606 | 56,473 |
| Ps⁻ arrived at the laser | 581,182 | 587,938 | 604,919 | 617,218 |
| Ps⁻ detached by the laser | 572,994 | 553,168 | 572,776 | 608,251 |
| Ps decayed in the laser | 169 | 181 | 180 | 189 |
| Laser photo-detachment efficiency | 91% | 94% | 95% | 99% |
| Ps which enter the Mach-Zehnder interferometer | 520,466 | 547,403 | 567,475 | 606,072 |
| Ps which have an angle with X axis less than 100 μrad | 6,448 | 4,252 | 22,477 | 8,943 |
| Ps which have an angle with Y axis less than 100 μrad | 6,569 | 4,194 | 22,795 | 9,422 |

**Table I.** Monte Carlo statistics for various simulation steps (see text).

Following the methodology proposed in Ref. 8, it is necessary to consider a very bright positron beam to perform interferometric measurements of Ps gravity on the time scale of the year. We propose to use a LINAC as primary electron beam to produce positron-electron pairs. For instance, the Milan BriXSinO facility proposes a superconductive LINAC operating at 92 MHz with an average electron current of 2.5 mA, an acceleration energy of 10 MeV [24-26]. At least $10^{16}$ fast electrons per second will be produced with this kind of LINAC with a beam spot lower than 1 mm and very narrow divergency of 50 μrad. These excellent characteristics allow us to predict a high e+/e- conversion efficiency, in particular using a neon-solid positron moderator a total conversion efficiency (cold e+)/(hot e-) of at least $10^{-6}$ is expected [27]. These characteristics can be compared with the performance of other LINAC-based positron sources in Ref. 27. The positron rate expectation of the slow beam is of around $10^{10}$ positrons per second with the technical conditions mentioned above. Considering a conversion efficiency of the target converter (tungsten decorated with sodium) for

Ps⁻/e+ is around $10^{-2}$ [7], our expectation is to obtain $10^8$ Ps⁻ per second emitted from the target. In contrast, using a LINAC as presented in Ref. 27 our expectation is to obtain $10^6$ Ps⁻ per second.

These simulations are conducted using a Gaussian Ps⁻ beam profile based on a standard positron beam dimension of 1 mm diameter (detailed in Sec. 2.2). However, an alternative approach, as described in Ref. 28, allows for remoderation of the primary positron beam, resulting in a Ps⁻ beam with a diameter one order of magnitude smaller [29]. Nevertheless, this option reduces the positron statistics of one order of magnitude. Despite the trade-off, the reshaping approach increases the positron density at the centre of the distribution, thereby improving the final Ps statistics relevant for Mach-Zehnder interferometry. Further investigation and a detailed study will be conducted to assess the feasibility and benefits of this approach.